\documentclass[showpacs,twocolumn,aps]{revtex4-1}
\usepackage[english]{babel}
\usepackage[utf8]{inputenc}
\usepackage{graphicx}
\usepackage{epstopdf}
\usepackage{amsmath, amssymb}
\usepackage{dcolumn}
\usepackage{subfigure}
\usepackage{amsmath,amsfonts,amsthm,bm}
\usepackage{mathtools}
\usepackage{float}

\DeclareMathOperator{\Tr}{Tr}

\newcommand{\bk}{{\bf{k}}}
\newcommand{\bp}{{\bf{p}}}
\newcommand{\qq}{{\bf{q}}}
\newcommand{\be}{\begin{equation}}
\newcommand{\ee}{\end{equation}}
\newcommand{\bq}{\begin{eqnarray}}
\newcommand{\eq}{\end{eqnarray}}

\begin{document}

\title{Interacting fermions in an expanding spacetime}

\author{L. N. Machado $^a$}\email[]{lucas.machado@fisica.unina.it}
\author{H. A. S. Costa $^b$}\email[]{helderfisica@gmail.com}
\author{I. G. da Paz $^b$}\email[]{irismarpaz@ufpi.edu.br}
\author{M. Sampaio $^c$}\email[]{marcos.sampaio@ufabc.edu.br}
\author{Jonas B. Araujo $^d$}\email[]{jonas.araujo88@gmail.com}

\affiliation{$^a$ Dipartimento di Fisica, Universit\`{a} degli Studi di Napoli ``Federico II", Via Cinthia, 21-80126, Napoli, Italy}

\affiliation{$^b$ Universidade Federal do Piau\'{\i}, Departamento de F\'{\i}sica, 64049-550, Teresina, PI, Brazil}

\affiliation{$^{c}$ Centre for Particle Theory, University of Durham, Durham, DH1 3LE, UK \\and CCNH, Universidade Federal do ABC,  09210-580 , Santo Andr\'e - SP, Brazil}

\affiliation{$^d$ Universidade Federal do Maranh\~ao, Centro de Ci\^{e}ncias Exatas e Tecnologia, 65080-040 , S\~ao Lu\'{\i}s, MA, Brazil}

\begin{abstract}
\noindent
We evaluate the effect of quantum electrodynamics  on  the  correlations between Dirac field modes corresponding  electron-positron pairs of opposite momenta generated by expansion of an asymptotically flat Friedmann-Robertson-Walker (FRW) universe.
The mutual information of out-going electron-positron pairs is evaluated 
to leading order in the coupling strength and compared with the free case. It is shown a decrease in the mutual information between the electron and positron. In addition, it is found that the change in the electron-positron mutual information depends on how the momentum is distributed between the positron and photon modes.
\end{abstract}

\maketitle
\section{Introduction}
Gravitational particle creation appeared as a fundamental process in early works on particle creation by black holes and black hole evaporation by S. Hawking in \cite{HAWKING7475}. One of the conclusions was that  gravitational collapse  converts  baryons and leptons in  the collapsing body  into entropy. The recent revival of the study of gravitational particle creation in a time-varying space-time background is partially motivated by the development of relativistic quantum information and  precise data acquired from cosmic microwave background radiation. These developments permit, for instance, to study quantum information processes in regimes where relativistic effects are important and to test fundamental predictions of inflation on primordial fluctuations, such as scale independence and gaussianity  \cite{PARKERCONF}.

Cosmological particle creation incorporates entropy production and particle back-reaction on space-time geometry \cite{LIMA}. In particular,  the degree of entanglement between quantum field modes has been shown to contain a fairly complete amount of information about the cosmic parameters characterizing the expansion \cite{MARTINEZ}. In \cite{BARANOV1}, the thermodynamic and kinetic properties of irreversible gravitationally induced particle production were studied in the context of Friedmann-Robertson-Walker (FRW) cosmology, showing that the resulting non-equilibrium distribution function has the same functional form of equilibrium, save that  the evolution laws should be corrected by the particle production process. Moreover, such a process was seen to affect predictions of the observable quantities as, for instance, the dark matter density and thermally averaged annihilation cross sections \cite{BARANOV2}.

In \cite{FUENTES0} the entanglement of momentum modes $(p,-p)$ of a free quantum scalar field, produced by an expanding conformally flat $1+1$-dimensional FRW universe, was shown to contain information about cosmic parameters characterising the space-time expansion. In the limit of small mass,  this allowed for expressing the cosmological parameters in terms of the amount of the entanglement, quantified by the von Neumann entropy ($S_{vN}$), generated throughout cosmic evolution. In a similar fashion,
in \cite{FUENTES1} it was studied the entanglement between
modes of opposite momenta of a free Dirac field. The latter
showed qualitative differences as compared with the
bosonic counterpart, namely that fermionic fields encode
more information about the underlying space-time than the
bosonic case. In other words, the response of the entanglement to the dynamical evolution of the universe was shown to depend on the nature of the quantum field. 
Whilst for the bosonic case $S_{vN}$ of an observed mode decreases
monotonically from a maximum at $p = 0$, for a Dirac field $S_{vN}$ peaks around a certain (optimal)
momentum $p > 0$. This shows a  privileged momentum range for which space-time expansion generates a large amount of entanglement for fermion fields as if the exclusion principle impeded entanglement between small momentum values in contrast to the bosonic case. Moreover,
the frequency at which this peak in $S_{vN}(|\bp|;m) $ occurs is sensible to the rapidity ($\rho$) of the expansion, while the amount of entanglement in the maximally entangled $|\bp|$ is sensitive to the total volume ($\epsilon$) of expansion. As these are the only cosmological parameters characterising the scale factor of their model, it follows that the information about all the parameters of the expansion is codified in this peak. In this ideal context, where the universe follows an (asymptotically flat) expansion governed by two parameters $\epsilon$ and $\rho$, namely the volume and rapidity,  in \cite{FUENTES1} it is designed a protocol to extract such cosmological parameters from the peaked behaviour in $|\bp|$ of the electron entropy (as opposed to the bosonic case, which monotonically decreases as $|\bp|$ increases), namely the rapidity of the expansion is encoded in the frequency of the maximally entangled mode whereas the volume of the expansion is codified in the amount of entanglement generated for this optimal mode.

The problem of whether the presence of interactions  stimulates
or prevents the production of particles by the gravitational field as compared with the
creation of free particles has  been solved long ago \cite{BIRRELFORD79,LEAHYUNRUH83,BIRRELDAVIES78}. In \cite{LOTZE1,LOTZE2}, it  was investigated the effect of quantum electrodynamics (QED) interaction  upon the creation of photons and electron-positron pairs both in a FRW expanding universe and in other expanding asymptotically minkowskian spacetimes in the distant past and future . The emphasis was on photon generation later than the Compton time of Dirac particles along the cosmological evolution because it is the mass of Dirac particles that allows the breaking of conformal symmetry. Therefore, cosmological creation of Dirac electron-positron pairs is  possible even in the absence of the QED interaction. On the other hand, the simultaneous creation of  electron-positron pairs and photons in a curved background with electromagnetic interaction can be studied within perturbation theory in $e^2/(4 \pi) \approx 1/137$  ($\hbar = c = 1$) as in \cite{LOTZE2} where it was evaluated an attenuation effect for fermion production. It would certainly be interesting to assess the role played by interactions in the quantum correlations over a time-varying cosmological background.

In \cite{HELDER1}, we studied  the  effect  of
interactions in the  evaluation of cosmological parameters through correlations of opposite momentum modes of the created particles in the simplest possible setup: a bosonic scalar field $\varphi$ subject to a $\lambda \varphi^4$ interaction immersed in an asymptotically flat $(1+1)$-dimensional
FRW expanding spacetime. The Bogolyubov coefficients were computed perturbatively in the interaction picture {\it{a la}} Birrel and Ford \cite{BIRRELFORD79} and served to conclude that 
self-interaction amplifies the low-frequency modes of the
scalar field producing an enhancement in the entanglement entropy between the particle pairs created by space-time expansion. Because there exist fundamental differences in entanglement generation between opposite momenta of the produced particles  due to their statistics, it is worth investigating the effect of interactions in the  entanglement and correlations between Dirac particles. This is the main purpose of this contribution which is organised as follows: in section II we present the quantisation of Dirac equation and the Maxwell field  in a conformally flat curved space-time. Section III contains all the main results of this work, namely the mutual information between electrons and positrons in both free and interacting cases, which are illustrated with the graphs in the end of this section. We chose the mutual information for it is a more meaningful quantity when dealing with $N$-partite ($N>2$) systems, since the interaction produces a third particle in the system, i.e. a photon, to leading order in perturbation theory. Moreover, it reduces to (twice) the von Neumann entropy in the free case limit as computed in \cite{FUENTES1}. Our concluding remarks are addressed in section IV and technical details are left to an appendix.

\section{The model} 

Consider the action of QED with Dirac fermions of mass $m$  embedded in a curved space-time background ${\cal{M}}$ in $(3+1)$-dimensions: 
\begin{eqnarray}
S_{QED}^{\cal{M}} &=& \int d^4x\sqrt{-g}[\frac{1}{2}\bar{\psi}(i\gamma^{\mu}(\partial_{\mu} - \Gamma_{\mu}) - m)\psi \nonumber \\ &-& \frac{1}{4}F_{\mu\nu}F^{\mu\nu} \nonumber  
-ie\bar{\psi}\gamma^{\mu} A_{\mu}\psi]. 
\label{action}
\end{eqnarray}   
We have adopted natural units. In $S_{QED}^{\cal{M}}$,  $e$ is the coupling constant, $F_{\mu\nu} = \nabla_{\mu}A_{\nu} - \nabla_{\nu}A_{\mu}$ is the electromagnetic field strength tensor, where $\nabla_{\mu}$ represents covariant differentiation, $\Gamma_{\mu}$ is the spinorial affine connection  defined as
\begin{equation}
\Gamma_{\mu} = -\frac{1}{8}[\gamma_a, \gamma_b]e^{a\nu}\nabla_{\mu}e^{b \nu},
\end{equation}    
in which $e_a^{\mu}$ is a tetrad field and $\gamma^{\mu} = e_a^{\mu}\gamma^a$ are the Dirac matrices in  curved spacetime, satisfying $\{ \gamma^\mu, \gamma^\nu\} = 2 g^{\mu \nu}$,   $\{ \gamma^a, \gamma^b\} = 2 \eta^{ab}$ and $e^a_\mu \, e^b_\nu \eta_{ab} = g_{\mu \nu}$.
We take a conformally flat FRW space-time with metric
\begin{equation} \label{Metric}
ds^2 = C(\eta)(-d\eta^2 + dx^i dx_i),
\end{equation}
where $\eta$ is the conformal time related to the cosmological time $t$ as $d\eta = dt/C(\eta)$ and we adopt the mostly plus signature.
Exact solutions in quantum field theory on curved space-times are notoriously difficult. In order to get analytical results, we follow \cite{FUENTES1} and \cite{DUNCAN} by choosing
\begin{equation}
C (\eta) = (1 + \epsilon(1 + \tanh(\rho\eta)))^2,
\label{scalefactor}
\end{equation}
where $\epsilon$, $\rho$ are positive real parameters which represent the total volume and the rapidity of the expansion, respectively. In the asymptotic past/future
\begin{equation}
C_{\mathrm{in}} \equiv C(\eta \rightarrow -\infty)=1, 
C_{\mathrm{out}} \equiv C(\eta \rightarrow +\infty)=(1+2 \epsilon)^2, 
\end{equation}
showing that space-time is asymptotically Minkowskian in ${\mathrm{in/out}}$-regions, respectively. In the asymptotic regions, space-time admits  time-like Killing vectors $ \pm \partial_{\eta}$. Consequently we may classify the solutions of the field equations  into positive and negative frequency modes and proceed with field quantization in terms of creation and annihilation operators. 

In order to study interacting fields, it is convenient to  work in the interaction picture. The field operators satisfy the free Dirac and Maxwell wave equations,
\begin{equation} \label{EqDirac}
[i\gamma^{\mu}(\partial_{\mu} - \Gamma_{\mu}) + m]\psi = 0,   
\end{equation}   
and
\begin{equation} \label{EqMaxwell}
\nabla^{\mu}\nabla_{\mu}A_{\nu} - \nabla_{\nu}(\nabla^{\mu}A_{\mu}) + R^{\alpha}_{\,\,\mu}A_{\alpha} = 0,
\end{equation}
where $R_{\mu\nu}$ is the Ricci tensor that arises due to the commutation relation for the covariant derivatives acting on a vector field. The state-vector of the system satisfies the Schroedinger equation
\begin{equation}
H_{\mathrm{int}}|\Psi\rangle = i\partial_{\eta}|\Psi\rangle,
\end{equation}   
with
\begin{equation} \label{Hi}
H_{\mathrm{int}} = -ie\int d^3x \sqrt{-g} \bar{\psi} \gamma^{\mu} A_{\mu} \psi.
\end{equation}

\subsection{Dirac field in an expanding spacetime}

Using $g_{\mu\nu}(x) = C(\eta)\eta_{\mu\nu}$, we can rewrite the Dirac equation (\ref{EqDirac}) by replacing the tetrad field with the square root of the scalar factor
\begin{equation}
[i\sqrt{C(\eta)}\gamma^{\mu}(\partial_{\mu} - \Gamma_{\mu}(C(\eta))) + m]\psi = 0.
\end{equation}
Spatial translational invariance of the metric (\ref{Metric}) permits us to factorize the solutions as
\begin{eqnarray} \label{Psi0}
\psi(x) &=& e^{i\bf{p}\cdot\bf{x}}[C(\eta)]^{-3/4} \\
&\times&(\gamma^{0}\partial_{\eta} + i{\bm{\gamma}}\cdot{\bf{p}} - m\sqrt{C(\eta)})\phi_p(\eta). 
\end{eqnarray}
Inserting this equation into the Dirac equation (\ref{EqDirac}), we obtain the following differential equation
\begin{equation}\label{EqDiracRW2}
\partial_{\eta}^2\phi_{p}^{(\pm)} + \left(m^{2}C(\eta) \pm i\frac{m\dot{C}(\eta)}{\sqrt{C(\eta)}} + p^2\right)\phi_{p}^{(\pm)} = 0
\end{equation}
where $p^2 = |{\bf{p}}|^2$, $\dot{C}(\eta) = \partial C(\eta)/\partial \eta$ and we have used that $\gamma^{0}$ has eigenvalues  $\pm 1$. Notice that (\ref{EqDiracRW2}) is just the harmonic oscillator equation with time-varying frequency
\begin{equation}
\omega_p^{2(\pm)}(\eta) = p^2 + m^{2}C(\eta)\pm i\frac{m\dot{C}(\eta)}{\sqrt{C(\eta)}}.
\label{eqnomegapm}
\end{equation} 
The solutions of Eq. (\ref{EqDiracRW2}) in the asymptotic regions are
\begin{equation} 
\label{Phi0}
\phi_{p}^{\mathrm{in}(\pm)}\Big|_{\eta \rightarrow -\infty}= \frac{e^{i\omega_{\mathrm{in}} \eta}}
{\sqrt{2\omega_{\mathrm{in}}^{(\pm)}}},\,
\phi_{p}^{\mathrm{out}(\pm)}\Big|_{\eta \rightarrow +\infty}= \frac{e^{-i\omega_{\mathrm{out}}\eta}}
{\sqrt{2\omega_{\mathrm{out}}^{(\pm)}}}
\end{equation}
with frequencies  $\omega_{\mathrm{in,out}} = \left(p^2 +\mu_{\mathrm{in,out}}^2\right)^{1/2}$ where $\mu_{\mathrm{in}}^2 = m^2 C(-\infty)$ and $\mu_{\mathrm{out}}^2 = m^2 C(+\infty)$ and we have used
$\dot{C}(\pm \infty) = 0$. Following the notation as in \cite{DUNCAN}, we represent the quantum field over the regions ``in" and ``out" with a pair of mode functions  $U^{\mathrm{in,out}}$, $V^{\mathrm{in,out}}$, where the $U$'s ($V$'s) are related to $\phi^{(-)}$ ($\phi^{(+)}$). Moreover, according to our conventions,  flat space spinors satisfy
\begin{equation}
\gamma^{0}u(0,s) = -u(0,s), \,\,
\gamma^{0}v(0,s) = v(0,s),
\end{equation}
with $s = 1, 2$ accounting for spin states. Putting all together, the curved-space spinor solutions of Dirac equation can be written in terms of the mode functions in the in-region,
\begin{eqnarray} \label{Uin}
U^{\mathrm{in}}_{\bp}(x; s ) &=& K_{\mathrm{in}}(p)[C(\eta)]^{-3/4}[-i\partial_{\eta} + i\bm{\gamma}\cdot\bf{p} \nonumber  \\ &-& m\sqrt{C(\eta)}]\phi^{\mathrm{in}(-)}_{p}e^{i\bf{p}\cdot\bf{x}}u(0,s), \nonumber \\
V^{\mathrm{in}}_{\bp}(x; s ) &=& K_{\mathrm{in}}(p)[C(\eta)]^{-3/4}[i\partial_{\eta} - i\bm{\gamma}\cdot\bf{p} \nonumber  \\ &-& m\sqrt{C(\eta)}]\phi^{\mathrm{in}(+)}_{p}e^{-i\bf{p}\cdot\bf{x}}v(0,s),
\label{eqnUV}
\end{eqnarray}
or out-region with similar expressions substituting $in$ with $out$, 
where 
\begin{equation}
K_{\mathrm{in}(\mathrm{out})}(p) = -\frac{1}{p}\left(\frac{\omega_{\mathrm{in}(\mathrm{out})} - \mu_{\mathrm{in}(\mathrm{out})}}{2\mu_{\mathrm{in} (\mathrm{out})}}\right)^{1/2}.
\end{equation}  
Finally the field operator can be expanded in terms of creation and annihilation operators in the in-region as 
\begin{equation}
\hat{\psi} (x) = \,\,\, {\mathclap{\displaystyle\int}\mathclap{\textstyle\sum}}_{\bp,s}
\left(\frac{\mu_{\mathrm{in}}}{\omega_{\mathrm{in}}}\right)^{\frac{1}{2}}
[\hat{a}^{\mathrm{in}}_{\bp, s}U^{\mathrm{in}}_p(x; s) + \hat{b}^{\dagger\mathrm{in}}_{\bp, s}V^{\mathrm{in}}_p(x; s )],
\label{eqnpsi}
\end{equation}
where $ \int_{\bp} \equiv \int d^3p/(2\pi)^{3/2}$ and similarly for the field operator in the out-region. The creation and annihilation operators which act on fermionic states satisfy the usual relations
\begin{equation}
\left\lbrace \hat{a}^{\mathrm{in (out)}}_{\bp, s}, \hat{a}^{\mathrm{in (out)}\dagger}_{\bp', s'} \right\rbrace = \left\lbrace \hat{b}^{\mathrm{in (out)}}_{\bp, s}, \hat{b}^{\mathrm{in (out)}\dagger}_{\bp', s'} \right\rbrace = \delta_{ss'}\delta_{\bp \bp'} 
\end{equation}
defining the in and out-vacua
\begin{equation}
\hat{a}^{\mathrm{in(out)}}_{\bp, s}\left|0\right\rangle_{\mathrm{in(out)}} = \hat{b}^{\mathrm{in(out)}}_{\bp, s}\left|0\right \rangle_{\mathrm{in(out)}} = 0.
\end{equation}
Bogolyubov's coefficients are defined as usual, connecting in- and out- regimes
\begin{equation}
\phi_p^{\mathrm{in} (\pm)}(\eta) = \alpha_p^{(\pm)}\phi_p^{\mathrm{out} (\pm)}(\eta) + \beta_p^{(\pm)}\phi_p^{\mathrm{out} (\mp) *}(\eta),
\end{equation}
and obey, according to our normalization, 
\begin{equation}
\alpha^{(-)}\alpha^{(+)*}-\beta^{(-)}\beta^{(+)*}= 1. 
\end{equation}
Equivalently, one can express the relations between the mode functions $U_{\mathrm{in/out}}, V_{\mathrm{in/out}}$ with the help of (\ref{eqnUV}):
\begin{eqnarray}
U^{\mathrm{in}}_\bp &=& \frac{K_{\mathrm{out}}}{K_{\mathrm{in}}}\left[\alpha_{p}^{(-)} U^{\mathrm{out}}_\bp + \beta_{p}^{(-)} V^{*\mathrm{out}}_\bp\right] \nonumber \\
V^{\mathrm{in}}_\bp &=& \frac{K_{\mathrm{out}}}{K_{\mathrm{in}}}\left[\alpha_{p}^{(+)}V^{\mathrm{out}}_\bp + \beta_{p}^{(+)}U^{*\mathrm{out}}_\bp\right].
\end{eqnarray}

A relation between  creation and annihilation operators in the asymptotic regimes can  be obtained from (\ref{eqnpsi}) using that the expansion in both regions describes the same field operator. We obtain
\begin{eqnarray}
\hat{a}^{\mathrm{out}}_{\bp,s} &=& {\cal{N}}
 \big[\alpha_{p}^{(-)}\hat{a}^{in}_{\bp,s} + \beta_{p}^{*(-)}\sum_{s'} X_{s s'}(-\bp)\hat{b}^{\dagger \, in}_{-\bp,s'}  \big], \nonumber \\
 &\equiv& \tilde{\alpha}_p \hat{a}^{\mathrm{in}}_{\bp,s}+ \tilde{\beta^*_p} \hat{b}^{\mathrm{\dagger in}}_{-\bp,s}\,\, , \nonumber \\
\hat{b}^{\mathrm{out}}_{\bp,s} &=&
{\cal{N}} \big[\alpha_{p}^{(-)}\hat{b}^{in}_{\bp,s} + \beta_{p}^{*(-)}\sum_{s'} X_{s s'}(-\bp)\hat{a}^{\dagger \, in}_{-\bp,s'}  \big] \nonumber \\
&\equiv&  \tilde{\alpha}_p \hat{b}^{\mathrm{in}}_{\bp,s}+ \tilde{\beta^*_p} \hat{a}^{\mathrm{\dagger in}}_{-\bp,s}\,\, , 
\label{eqn:bogo}
\end{eqnarray}
with 
\begin{equation}
{\cal{N}} = \Big[ \frac{\omega_{\mathrm{out}}}{\omega_{\mathrm{in}}} \frac{\omega_{\mathrm{in}} - C_{\mathrm{in}}^{1/2}}{\omega_{\mathrm{out}}-C_{\mathrm{out}}^{1/2}}\Big]^{1/2}
\end{equation}
and the so called polarization tensor  $X_{s s'}(p)$ is   given by \cite{DUNCAN}
\begin{equation}
X_{s s'}(\bp) = - 2C_{\mathrm{out}}  \left (\frac{\omega_{\mathrm{out}}- C_{\mathrm{out}}^{1/2}}{2 p^2 C_{\mathrm{out}}} \right)^{1/2} U^{\mathrm{out}}_{-\bp} (s') v (0,s),
\end{equation}
satisfying 
\bq
\sum_{s'} |X_{ss'}(\bp)|^2 &=& 2 \mu_{\mathrm{in(out)}}K_{\mathrm{in(out)}}^2
(\omega_{\mathrm{in(out)}}-\mu_{\mathrm{in(out)}})\delta_{ss'}\nonumber \\&=& \sum_{s'}
\Big[\frac{\mu_{\mathrm{in(out)}}}{p} \Big( 1 - \frac{\omega_{\mathrm{in(out)}}}{\mu_{\mathrm{in(out)}}}\Big)\Big]^2 \delta_{ss'}.
\label{gamma1}
\eq

In a similar fashion, we may obtain expressions for $\hat{a}^{\dagger \mathrm{out}}_{\bp,s}$ and $\hat{b}^{\dagger \mathrm{out}}_{\bp,s}$. 

The Bogolyubov coefficients can be analytically evaluated for the asymptotically free space-time (\ref{Metric}) using the solutions of equation (\ref{EqDiracRW2}) \cite{FUENTES0,FUENTES1,HELDER1,DUNCAN}. They read:
\begin{eqnarray}
\alpha_{p}^{(\pm)} & = &\left (\frac{\omega_{\mathrm{in}}}{\omega_{\mathrm{out}}}  \right )^{1/2} \frac{\Gamma \Big(1-\frac{i\omega_{\mathrm{out}}}{\rho}\Big)}{\Gamma\Big(\frac{-i}{2\rho}(\omega_- \pm i\epsilon)\Big)} \times \nonumber \\ 
&\times& \frac{\Gamma \Big(\frac{i\omega_{\mathrm{in}}}{\rho}\Big)}{\Gamma \Big(1- \frac{i}{2 \rho} (\omega_- \pm i \epsilon)\Big)}
\label{gamma2}
\end{eqnarray}
and
\begin{eqnarray}
\beta_{p}^{(\pm)} & = & -\left (\frac{\omega_{\mathrm{in}}}{\omega_{\mathrm{out}}}  \right )^{1/2} \frac{\Gamma \Big(1-\frac{i\omega_{\mathrm{out}}}{\rho}\Big)}{\Gamma\Big(\frac{-i}{2\rho}(\omega_+\mp i\epsilon)\Big)} \times \nonumber \\ 
&\times& \frac{\Gamma \Big(\frac{-i\omega_{\mathrm{in}}}{\rho}\Big)}{\Gamma \Big(1- \frac{i}{2 \rho} (\omega_+ \mp i \epsilon)\Big)},
\label{gamma3}
\end{eqnarray}
where we used (\ref{eqnomegapm}) to conclude that
\bq
\omega_{\mathrm{in}}^\pm - \omega_{\mathrm{out}}^\pm &=& \omega_{\mathrm{in}} - \omega_{\mathrm{out}} \pm i \epsilon\nonumber \\
&\equiv& \omega_- \pm i \epsilon ,
\eq
and 
\bq
\omega_{\mathrm{in}}^\pm + \omega_{\mathrm{out}}^\pm &=& \omega_{\mathrm{in}} + \omega_{\mathrm{out}} \mp i \epsilon\nonumber \\
&\equiv& \omega_+ \mp i \epsilon.
\eq

\subsection{Electromagnetic field in an expanding spacetime}

As it is well known, the free photon Lagrangian is invariant  under local rescalings in the metric and so is the Dirac Lagrangian for massless fermions in four dimensions. In fact, the addition of a QED interaction term does not violate conformal invariance either. That means that particle creation mechanism for electrons, positrons and photons depend on the mass term in the Dirac field. Concretely one may perform a transformation $g'_{\mu\nu} = \Omega^2 g_{\mu\nu}$ and $A^{'\mu} = \Omega^{-2} A^{\mu}$  so that $g'_{\mu\nu} = \eta_{\mu\nu}$. Hence ``in" and ``out" solutions will be undistinguishable and the vector potential $\hat{A}_{\mu}$ may be expanded in terms of plane-wave modes in Minkowski space  
\begin{equation} 
\hat{A}_{\mu} (x) = {\mathclap{\displaystyle\int}\mathclap{\textstyle\sum}}_{k,\sigma} \frac{1}{\sqrt{2k}}\epsilon_{\mu}^{\sigma}(k)
\left(\hat{c}_{\bk, \sigma} e^{-i {\bf{k}}\cdot {\bf{x}}} + \hat{c}_{\bk, \sigma}^{\dagger}e^{i{\bf{k}}\cdot {\bf{x}}} \right),
\label{eqnA}
\end{equation}
where $\epsilon_{\mu}^{\sigma}$ is the polarization vector for the polarization state $\sigma$ and $\hat{{c}}_{\bk, \sigma}$ and $\hat{{c}}_{\bk, \sigma}^{\dagger}$  are the annihilation and creation   operators for photons, respectively. These operators satisfy the usual commutation relations
\begin{align}
\begin{split}
[\hat{{c}}_{\bk, \sigma}, \hat{{c}}_{\bk', \sigma'}^{\dagger}] &= \delta_{\sigma, \sigma'}\delta_{\bk, \bk'}, \\
[\hat{{c}}_{\bk, \sigma}, \hat{{c}}_{\bk', \sigma'}] &= [\hat{{c}}_{\bk, \sigma}^{\dagger}, \hat{{c}}_{\bk', \sigma'}^{\dagger}] = 0.
\end{split}
\end{align}
The natural choice for the vacuum state is conformal vacuum of the Maxwell theory and it is defined as $\hat{{c}}_{k, \sigma}|0\rangle = 0$.

\section{Particle Creation, entanglement and mutual information} 

In \cite{FUENTES0,FUENTES1} it was shown that the vacuum of the quantum field in the asymptotic past evolves to an entangled state in the      asymptotic future. Moreover the entanglement generated by the expansion was shown to contain information about the cosmic evolution, such an information being more easily obtained for fermionic  as compared with bosonic fields. The entanglement entropy as a function of  physical parameters such as momentum and mass or the cosmological parameters $\epsilon$ and $\rho$ was qualitatively distinct for bosons or fermions. In \cite{LOTZE1,LOTZE2} it was estimated the magnitude of the contributions to particle
creation, which are due to the interaction in comparison with the creation of free
electron-positron pairs. In particular,  it was verified an attenuation effect for fermionic particles as free electron-positron pairs counteract pair creation due to non-gravitational interaction. Effectively this means that the electromagnetic interaction produces less than one electron-positron  pair per photon to leading order. 

\begin{figure}[h]
	\centering
	\includegraphics[scale=0.20]{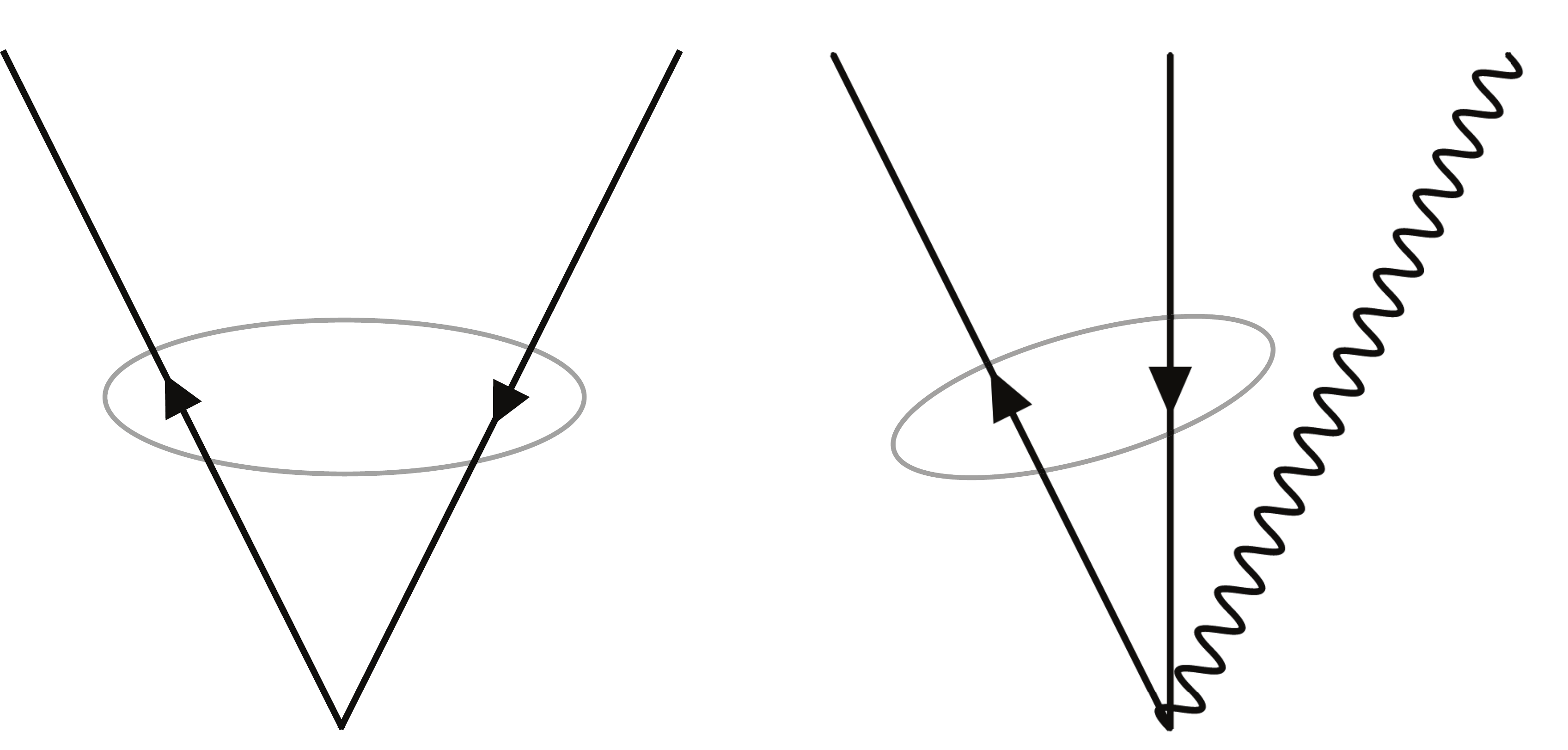}
	\caption{Creation of electron-positron pairs and photons  in an expanding space-time. Even in the absence of interactions, electron-positron pairs are created and realised as an entangled state in the modes ${\bf{p}}$,$-{\bf{p}}$. Electromagnetic interaction is expected to affect the mutual information of the electron-positron pair. }
	\label{figA}
\end{figure}

Here we are interested in the contribution of the electromagnetic interaction to the correlations generated between Dirac modes in an expanding conformally flat and asymptotically Minkowskian space-time.  Up to tree-level, the interaction term (\ref{Hi}) in an expanding cosmological background, where conservation of field energy is not required, can be depicted  as in fig. \ref{figA}. To evaluate such a contribution it is convenient to use $S$-matrix techniques in the interaction picture. We  assume that the electromagnetic interaction is adiabatically switched off in the ``in" and ``out" regions of spacetime. 

Let us formally write the state vector in the remote past as 
\begin{equation}
|\Psi\rangle_{\mathrm{in}} = N \Big( |0\rangle^{D}_{\mathrm{in}} \otimes |0\rangle^{\gamma}_{\mathrm{in}} + \int_{\bk_1, \bk_2, \bk_3}\Gamma_{\bk_1,\bk_2,\bk_3}|\bk_1, \bk_2\rangle \otimes |\bk_3\rangle \Big),
\label{instate}
\end{equation}
to leading order in the interaction, where $|0\rangle^{D}$ and $|0\rangle^{\gamma}$ correspond to the Dirac and photon vacua and ${\bf{k}}_1$ , ${\bf{k}}_2$ and ${\bf{k}}_3$ represent the electron, positron and photon momenta, respectively. Moreover, to guarantee that $|\Psi\rangle_{\mathrm{in}}$ is normalized to unity, we require
\be
|N|^{-2} = 1 + \int_{\bk_1, \bk_2, \bk_3} |\Gamma_{\bk_1,\bk_2,\bk_3}|^2,
\ee
where
\begin{align} \label{Gamma1230}
\begin{split}
\Gamma_{\bk_1,\bk_2,\bk_3} &=  \langle \bk_3| \otimes \langle  \bk_2, \bk_1| \Big( -i\int_{-\infty}^{\infty}\hat{H}_{\mathrm{int}} \,  d\eta  \Big) \, |0 \rangle^{D}_{\mathrm{in}} \otimes  | 0 \rangle^{\gamma}_{\mathrm{in}}, \\
& \equiv  e \, \delta^3 (\bk_1 + \bk_2 + \bk_3) A(\bk_1, \bk_2, \bk_3),
\end{split}
\end{align}
and $\hat{H}_{\mathrm{int}}$ is given by (\ref{Hi}).

\subsection{The non-interacting case}

In the interaction picture the Bogolyubov coefficients carry information only about the non-interacting contribution to the total particle creation. Due to the structure of the Bogolyubov transformation (\ref{eqn:bogo}), an electron-positron pair of modes $({\bf{p}},-{\bf{p}})$ can be produced by the free Dirac field in an expanding space-time. Moreover, a Fock space state in the in-region is related to the corresponding state in the out-region by a Schmidt decomposition which can be written as  a two-mode squeezing unitary operator factorised for each pair of modes. Formally,
\bq
|0\rangle_{\mathrm{out}}^D &=& \prod_{\bp} |0_{\bp} 0_{-\bp}\rangle_{\mathrm{out}} \,\, , \nonumber \\
|0_{\bp} 0_{-\bp}\rangle_{\mathrm{out}} &=& {\cal{S}}(\hat{a}_{\bp}^{\mathrm{in}\dagger},\hat{b}_{-\bp}^{\mathrm{in}\dagger})|0_{\bp} 0_{-\bp}\rangle_{\mathrm{in}}
\label{ent-state} \\
&\equiv& \sum_{n = 0}^1 a_n |n_{\bp}n_{-\bp}\rangle, \nonumber
\eq
with 
\bq
{\cal{S}} &=&
 C_{p}\, \mathrm{exp}\Bigg[ {-\frac{\beta_{p}^{*(-)}}{{\alpha}_{p}^{(-)}} \sum_{s'}X_{ss'}(-\bk)\hat{a}_{\bp,s}^{\mathrm{in}\dagger}\hat{b}_{-\bp,s'}^{\mathrm{in} \dagger}} +\mathrm{h.c.} \Bigg]\nonumber \\
 a_n &=&  (-1)^n C_{p} \left[\frac{\beta_{p}^{*(-)}}{\alpha_{p}^{(-)}} \sum_{s'} X_{s s'}(-\bold{p})\right]^n.
\eq
and $C_{p}$ is a normalisation constant to be determined so that $^{D}_{out}\langle 0|0\rangle_{\mathrm{out}}^D = 1$ to yield
\be
C_{p} =  ( 1 + \gamma_p)^{-1/2},
\ee
where
\be
\gamma_p \equiv \sum_{s'} \Bigg|\frac{\beta_p^{*(-)}}{\alpha_p^{(-)}}\Bigg|^2 |X_{s s'}|^2 ,
\label{gammafermion}
\ee
which can be computed  for a FRW conformally flat space-time of scale factor (\ref{scalefactor}) with the help of equation (\ref{gamma1}), (\ref{gamma2}) and (\ref{gamma3}) to yield \cite{MARTINEZ}:
\bq
\gamma_p &=&\frac{(\omega_- +m\epsilon)(\omega_+ +
	m\epsilon)}{( \omega_-
	- m\epsilon)(\omega_+ -m\epsilon)} \times \nonumber \\
&\times&  \frac{\sinh\left[\frac{\pi}{\rho} (\omega_- -
	m\epsilon)\right]\sinh\left[\frac{\pi}{\rho} (\omega_- + m\epsilon)\right]}{\sinh\left[\frac{\pi}{\rho}(\omega_+ +m\epsilon)\right]\sinh\left[\frac{\pi}{\rho}(\omega_+ -m\epsilon)\right]}\times \nonumber \\
&\times& \Big[\frac{m}{p} \Big( 1 - \frac{\omega_{\mathrm{in}}}{m}\Big)\Big]^2.
\label{gammaexplicit}
\eq
Expression  (\ref{ent-state}) means  that the pure, entangled state $|0_{\bp} 0_{-\bp}\rangle_{\mathrm{out}}$ can be written as a Schmidt decomposition in all number states $n$ in the bi-partition $\bp, -\bp$, where $n_{\bp} (n_{-\bp})$ labels the number of electron (positron) excitations in the field mode $\bp$ as seen by an observer (inertial) in the $in$-region. A similar decomposition can be made for  $|0_{\bp} 0_{-\bp}\rangle_{\mathrm{in}}$, the interpretation being that particles are created by cosmic expansion. In order to quantify the entanglement of particle-antiparticle modes in $|0_{\bp} 0_{-\bp}\rangle_{\mathrm{out}}$, we define the density matrix,
\be
\varrho_{\bp,-\bp}^0 = |0_{\bp} 0_{-\bp}\rangle_{\mathrm{out}} \langle 0_{\bp} 0_{-\bp} |,
\ee
from which the reduced density matrix for the electron reads
\bq
\varrho_{\bp}^0&=& \sum_n \langle n_{-\bp}|\varrho_{\bp,-\bp}^0|n_{-\bp}\rangle\nonumber \\
&=&\frac{\sum_{n=0}^1 \big(\gamma_p\big)^n |n_{\bp} \rangle_{\mathrm{in}}\langle n_{\bp}|}{(1+\gamma_p)}.
\label{Order0}
\eq
The von Neumann entropy $\mathbb{S}$ of the reduced density matrix $\varrho_{\bp}$,
\bq
\mathbb{S}_{e^-}^0 &=& -\mathrm{tr} \big(\varrho_{\bp}^0 \mathrm{log}_2 \varrho_{\bp}^0\big)\nonumber \\
&=& \mathrm{log}_2\Bigg( \frac{1+ \gamma_p}{\gamma_p\,\,^{\gamma_p/(1+ \gamma_p)}}\Bigg).
\eq
is a well-founded measure of entanglement between electrons and positrons produced in modes $p$, $-p$. From (\ref{gammaexplicit})
one sees that the entropy is zero when the mass of the fermion field vanishes.

It is important to notice that the density of created particles is given by  $|\beta_\bp^*|^2$, and in order to the total number of particles in all modes be finite, namely $\int n_{\bp} d^3 \bp<\infty$,  $|\beta_\bp^*|^2$ should fall faster than $|\bp|^{-3}$ as $|\bp| \rightarrow \infty$. Such a condition also validates the normalisation of the $out$-vacuum expressed in terms of the $in$-vacuum (squeezing) \cite{MU}.
In the particular space-time we consider  \cite{FUENTES0,FUENTES1}, this condition is met. In particular, regarding the entropy calculation in the free case, for both bosons and fermions, it is possible to analytically evaluate the entanglement at infinity. For instance, for $m=|\bp|=\rho=1$, it is found that in the limit where $\epsilon \rightarrow \infty$,
\begin{equation}
\gamma_p\rightarrow e^{-\pi \sqrt{2}} \frac{e^{\pi \sqrt{2}}-e^{\pi}}{e^{\pi \sqrt{2}+1}-1} \ ,\  \mathbb{S}_{e^-}^0(\epsilon \rightarrow \infty) \approx 0.0048,
\end{equation}
where $m$ is the mass, $\epsilon$ is the total volume and $\rho$ the rapidity of the expansion. For purposes of comparison with the interacting case, we shall compute the mutual information between the electron-positron in the free case, which is defined as
\begin{eqnarray}
\mathcal{I}_{e^-e^+}^0 &=& \mathbb{S}_{e^-}^0 + \mathbb{S}_{e^+}^0 - \mathbb{S}_{e^-e^+}^0\nonumber\\
&=&2\,\mathbb{S}_{e^-}^0,\label{Minfo-free}
\end{eqnarray}
because $\mathbb{S}_{e^-e^+}^0=0$, as the system is a pure state, and $\mathbb{S}_{e^+}^0=\mathbb{S}_{e^-}^0$.
\subsection{Correlation measure of the dressed vacuum}

In order to study the effect of the electromagnetic in- teraction on the entropy generation for electrons of mode $\bp$ we apply the squeezing operator in the fermionic sector in (\ref{instate}), while leaving the (massless) photonic sector unaltered $|\Psi\rangle_{\mathrm{out}}=({\cal{S}}\otimes\hat{1}) |\Psi\rangle_{\mathrm{in}}$, namely 
\begin{equation}
|\Psi\rangle_{\mathrm{out}} = N \Big( {\cal{S}}|0\rangle^{D}_{\mathrm{in}} \otimes |0\rangle^{\gamma}_{\mathrm{in}} + \int_{\bk_1, \bk_2, \bk_3} \hskip-0.5cm\Gamma_{\bk_1,\bk_2,\bk_3} {\cal{S}}|\bk_1, \bk_2\rangle \otimes |\bk_3\rangle \Big).
\label{outstate}
\end{equation}
The integrand in the RHS of the expression above can be evaluated as
\bq
&&{\cal{S}}|\bk_1, \bk_2\rangle \otimes |\bk_3\rangle \longrightarrow  \sqrt{2 \omega_1} \sqrt{2 \omega_2} \sqrt{2 k_3} \nonumber \\ && {\cal{S}}\hat{a}^{\mathrm{in}\dagger}_{\bk_1, s}{\cal{S}}^{-1}{\cal{S}}\hat{b}^{\mathrm{in}\dagger}_{\bk_2, s'}{\cal{S}}^{-1}{\cal{S}}|0\rangle^D_{\mathrm{in}}  \otimes \hat{{c}}^{\mathrm{in}\dagger}_{\bk_3, \sigma}|0\rangle^{\gamma},
\eq
and hereafter we omit the index  ``in" at the creation operator for photons because of the conformal invariance of Maxwell theory. Moreover 
\bq
{\cal{S}}\hat{a}^{\mathrm{in}\dagger}_{\bk_1, s}{\cal{S}}^{-1} &=& \hat{a}^{\mathrm{out}\dagger}_{\bk_1, s}\,\, , \nonumber \\ {\cal{S}}\hat{b}^{\mathrm{in}\dagger}_{\bk_2, s}{\cal{S}}^{-1} &=& \hat{b}^{\mathrm{out}\dagger}_{\bk_2, s} \nonumber \\ {\mathrm{and}} \quad {\cal{S}}|0\rangle^D_{\mathrm{in}} &=& |0\rangle^D_\mathrm{out}. \nonumber
\eq
In order to evaluate the contribution of the electromagnetic interaction to leading order, we write the out-state as
\begin{equation} \label{Psiout}
| \Psi \rangle_{\mathrm{out}} = N (\hat{1} + \hat{I}) (|0 \rangle^D_{\mathrm{out}} \otimes | 0 \rangle^{\gamma}),
\end{equation}
where
\begin{equation} \label{I}
\hat{I} = \int_{\bk_1, \bk_2, \bk_3} \Gamma_{\bk_1,\bk_2,\bk_3} \left(\hat{a}^{\mathrm{out}\dagger}_{\bk_1, s}\hat{b}^{\mathrm{out}\dagger}_{\bk_2, s}\hat{{c}}^{\dagger}_{\bk_3, \sigma}\right).
\end{equation}
The dressed density matrix of the system reads
\bq
\tilde{\varrho} &=& |\Psi\rangle_{\mathrm{out}}\langle\Psi|  =  (\tilde{\varrho}^0 + \hat{I}\tilde{\varrho}^0 + \tilde{\varrho}^0 \hat{I}^\dagger +  \hat{I}\tilde{\varrho}^0  \hat{I}^\dagger ), \label{dressedrho}  \\
\tilde{\varrho}^0 &=& |N|^2 \,\, |0 \rangle^D_{\mathrm{out}}\langle 0| \otimes |0\rangle^\gamma \langle 0| \label{DMfull}.
\eq
Because now we are dealing with a tri-partite system, the von Neumann entropy of the reduced system is no longer a meaningful quantity. Rather, we employ the mutual information of the electron-positron pair as a consistent measure of correlation in both free and interacting cases. Mutual information has been used in \cite{PA,AN}, where it was employed the so-called average sub-system entropy in the context of multipartite systems.

In order to calculate the mutual information between the electron and positron with the dressed vacuum, we need to evaluate the reduced density of the electron, the positron and the system electron-positron, just so we can compute their individual von Neumann entropies. The mutual information in the interacting case reads
\begin{equation}
\tilde{\mathcal{I}}_{e^- e^+} = \tilde{\mathbb{S}}_{e^-} + \tilde{\mathbb{S}}_{e^+} - \tilde{\mathbb{S}}_{e^- e^+}.
\end{equation}

To compute the leading order  correction for outgoing electrons of momentum $\bp$, we must choose the partition  $\bk_1 = \bp$, $\bk_2 = a \bp + {\bf{q}} \equiv {\bf{q}}_1$ and $\bk_3 = -(1+a)\bp - {\bf{q}} \equiv \qq_2$, for any real $a$ and $\bp \cdot {\bf{q}} =0$, on momentum conservation grounds. Hence the reduced matrix for the electron is
\bq
\tilde{\varrho}_{\bp} &=& \sum_{j,l} 
\langle j_{\qq_1}|\otimes \langle l_{\qq_2}| \,\,  \tilde{\varrho} \,\, |j_{\qq_1}\rangle \otimes | l_{\qq_2}\rangle  \nonumber \\
&=& \sum_{j} 
\langle j_{\qq_1}|  \tilde{\varrho}^0  |j_{\qq_1}\rangle 
\nonumber \\
&+&  \sum_{j,l} \Big\{\langle j_{\qq_1},l_{\qq_2}|  (\hat{I}\tilde{\varrho}^0 + \tilde{\varrho}^0 \hat{I}^\dagger) |j_{\qq_1}, l_{\qq_2}\rangle  
\nonumber \\
&+& \langle j_{\qq_1},l_{\qq_2}| \hat{I}\tilde{\varrho}^0  \hat{I}^\dagger  |j_{\qq_1}, l_{\qq_2}\rangle \Big\} ,
\label{RDM-D}
\eq
where we have abbreviated the notation and  used equation (\ref{dressedrho}). To compute the effect of the dynamical space-time in the interaction picture, we write (\ref{I}) with help of (\ref{eqn:bogo}). According to the partition we chose, we have
\bq
\hat{I} &=& \Gamma_{\bp,\qq_1,\qq_2}\,  \hat{a}^{\mathrm{out}\dagger}_{\bp, s}\hat{b}^{\mathrm{out}\dagger}_{\qq_1, s}\otimes \hat{{c}}^{\dagger}_{\qq_2, \sigma} = \nonumber \\
&=& \Gamma_{\bp,\qq_1,\qq_2} \, \Big( 
\tilde{\alpha}_{\bp}^{*}\tilde{\alpha}_{\qq_1}^{*}\hat{a}_{\bp, s}^{\mathrm{in}\dagger}\hat{b}_{\qq_1, s}^{\mathrm{in}\dagger}  \nonumber \\ &+& \tilde{\alpha}_{\bp}^{*}\tilde{\beta}_{\qq_1}\hat{a}_{\bp, s}^{\mathrm{in}\dagger} \hat{a}_{-\qq_1, s}^{\mathrm{in}} + \tilde{\beta}_{\bp}\tilde{\alpha}_{\qq_1}^{*}\hat{b}_{-\bp, s}^{\mathrm{in}}\hat{b}_{\qq_1, s}^{\mathrm{in}\dagger} 
\nonumber\\
&+& \tilde{\beta}_{\bp}\tilde{\beta}_{\qq_1}\hat{b}_{-\bp,s}^{\mathrm{in}}\hat{a}_{-\qq_1, s}^{\mathrm{in}}
\Big) \otimes \hat{c}_{\qq_2,\sigma}^\dagger \nonumber \\ &\equiv& \Gamma_{\bp,\qq_1,\qq_2}\,\,  \hat{I}^D \otimes \hat{c}_{\qq_2,\sigma}^\dagger ,
\label{Ipartition}
\eq
where $\hat{I}^D$ acts on the Dirac sector only. The terms of the electron's reduced density matrix in (\ref{RDM-D}) are straightforwardly evaluated. The ${\cal{O}}(e^0)$  term describing a free Dirac particle reads:
\be
\sum_{j} 
\langle j_{\qq_1} |  \tilde{\varrho}^0 |j_{\qq_1}\rangle = \varrho_{\bp}^0,
\ee
as in (\ref{Order0}). The first two terms of ${\cal{O}}(e)$ in the double-sum in (\ref{RDM-D}) do not contribute since the partial trace over the photon space yields zero. As for the term of ${\cal{O}}(e^2)$, it can formally be written as:
\bq
&&\sum_{j,l} \langle j_{\qq_1},l_{\qq_2}| \hat{I}\tilde{\varrho}^0 \hat{I}^\dagger  |j_{\qq_1}, l_{\qq_2}\rangle =\nonumber \\
&& \sum_{j,l,n,n'} a_n a^{*}_{n'} |\Gamma_{\bp,\qq_1,\qq_2}|^2 \langle l_{\qq_2}|1_{\qq_2}\rangle \langle j_{\qq_1}|\hat{I}^D|n_{\bp}\rangle|n_{-\bp}\rangle \nonumber \times \\ && \times \, \langle 1_{\qq_2}|l_{\qq_2}\rangle \langle n'_{\bp}|\langle n'_{-\bp}|\hat{I}^{D\,\dagger}|j_{\qq_1} \rangle \, .
\label{formalrho2}
\eq
The term involving the operator $\hat{I}^D$ can be explicitly written as 
\bq
\langle j_{\qq_1}|\hat{I}^D|n_{\bp}\rangle|n_{-\bp}\rangle &=& \tilde{\alpha}_{\bp}^{*}\tilde{\alpha}_{\qq_1}^{*}
\large( \hat{a}_{\bp, s}^{\mathrm{in}\dagger}|n_{\bp}\rangle \large)\, \langle j_{\qq_1}| \hat{b}_{\qq_1, s}^{\mathrm{in}\dagger}|n_{-\bp}\rangle  +\nonumber \\
&+& \tilde{\alpha}_{\bp}^{*}\tilde{\beta}_{\qq_1} \large( \hat{a}_{\bp, s}^{\mathrm{in}\dagger} \hat{a}_{-\qq_1, s}^{\mathrm{in}}|n_{\bp}\rangle \large) \langle j_{\qq_1}|n_{-\bp}\rangle + \nonumber \\ &+& \tilde{\beta}_{\bp}\tilde{\alpha}_{\qq_1}^{*}  |n_{\bp}\rangle \,\, \langle j_{\qq_1}|\hat{b}_{-\bp, s}^{\mathrm{in}}\hat{b}_{\qq_1, s}^{\mathrm{in}\dagger}|n_{-\bp}\rangle +\nonumber \\
&+&
\tilde{\beta}_{\bp}\tilde{\beta}_{\qq_1}  \large(\hat{a}_{-\qq_1, s}^{\mathrm{in}}|n_{\bp}\rangle \large) \langle  j_{\qq_1}| \hat{b}_{-\bp,s}^{\mathrm{in}} |n_{-\bp}\rangle \nonumber .
\eq
It becomes  clear from the structure above that only the third term in the RHS contributes, which multiplied by its hermitian conjugate as shown in (\ref{formalrho2}), gives the following formal expression for the electron's reduced density matrix:
\bq
\tilde{\varrho}_{\bp} &=&N_{A} \,\big[ \varrho_{\bp}^0 + e^2 |a_1|^2|A (\bp,\qq_1,\qq_2)|^2 |\tilde{\beta}_{\bp}|^2 |\tilde{\alpha}_{\qq_1}^{*}|^2|1_{\bp,s}\rangle\langle1_{\bp,s}| \nonumber \\ &+& {\cal{O}}(e^4) \big],
\label{RDMF}
\eq
where $N_A$ is chosen so that $\Tr(\tilde{\varrho}_{\bp})=1$, and we used (\ref{Gamma1230}). Moreover, 
\bq
&&|A(\bp, \qq_1, \qq_2)|^2 = 
\frac{2 \epsilon^2 (1+\epsilon)^2}{\pi \rho^2 \sinh^2\Large(\frac{\pi \omega_{\mathrm{in}}(p)}{\rho}\Large)} \frac{[(1+a)^2 + q^2]}{p^2 \,\, m^2} \times \nonumber \\
&& \times \,\,\omega_{\mathrm{in}}(p) \omega_{\mathrm{in}}(q_1) [\omega_{\mathrm{in}}(p)-m]^3
[\omega_{\mathrm{in}}(q_1)-m]^3, 
\eq
which can be inferred from the explicit expression of $A(\bp, \qq_1, \qq_2)$ demonstrated in the appendix. It is noteworthy that other diagrams of ${\cal{O}}(e^2)$ yield zero and thus (\ref{RDMF}) is the total contribution to the  reduced density matrix for an electron of outgoing momentum $\bp$.

Finally, the von Neumann entropy of the reduced density matrix taking into account the electromagnetic interaction to ${\cal{O}}(e^2)$ can be  written in terms of its eigenvalues $\lambda_i$,%
\bq
\mathbb{\tilde{S}}_{e^-} &=& -\sum_{i=0}^{1} \lambda_i \log_2 \lambda_i, \label{S-elec}
\eq with
\bq
\lambda_0=\frac{|a_0|^2}{1+ e^2\,|a_1|^2|{A}|^2|\tilde{\beta}_{\bp}|^2 |\tilde{\alpha}_{\qq_1}^{*}|^2},
\eq
\bq
\lambda_1=\frac{|a_1|^2\big(1+e^2|{\cal{A}}|^2|\tilde{\beta}_{\bp}|^2 |\tilde{\alpha}_{\qq_1}^{*}|^2\big)}{1+ e^2\,|a_1|^2|{A}|^2|\tilde{\beta}_{\bp}|^2 |\tilde{\alpha}_{\qq_1}^{*}|^2},
\eq%
where $A=A (\bp,\qq_1,\qq_2)$.

In a similar fashion, by tracing out the electron and photon modes, we compute the positron's reduced density matrix 
\bq
\tilde{\varrho}_{\qq_1} &=&N_{B} \,\big[ \varrho_{-\bp}^0 + e^2 |A|^2 \big| \tilde{\alpha}_{\bp}^{*}\tilde{\alpha}_{\qq_1}^{*}a_0+\tilde{\beta}_{\bp} \tilde{\alpha}_{\qq_1}^{*}a^*_1\big|^2 \times \nonumber \\ &\times&  |1_{\qq_1,s}\rangle\langle1_{\qq_1,s}| + {\cal{O}}(e^4)\big],
\label{RDMP}
\eq
with emerging positrons  of  momenta: $-\bp$ for the free case and a contribution in $\qq_1$ due to the interaction. In addition, $N_B$ is a normalisation constant and $\varrho_{-\bp}^0$ is the positron's reduced density matrix in the free case, which has the same eigenvalues as the electron's. That said, the positron's entropy becomes%
\bq
\mathbb{\tilde{S}}_{e^+} &=& -\sum_{i=0}^{1} \lambda'_i \log_2 \lambda'_i, \label{S-pos}
\eq with
\bq
\lambda'_0=\frac{|a_0|^2}{1+ e^2 |{A}|^2 \big| \tilde{\alpha}_{\bp}^{*}\tilde{\alpha}_{\qq_1}^{*}a_0+\tilde{\beta}_{\bp} \tilde{\alpha}_{\qq_1}^{*}a^*_1\big|^2},
\eq
\bq
\lambda'_1=\frac{|a_1|^2}{1+ e^2 |{A}|^2 \big| \tilde{\alpha}_{\bp}^{*}\tilde{\alpha}_{\qq_1}^{*}a_0+\tilde{\beta}_{\bp} \tilde{\alpha}_{\qq_1}^{*}a^*_1\big|^2},
\eq%
\bq
\lambda'_2=\frac{e^2 |{A}|^2 \big| \tilde{\alpha}_{\bp}^{*}\tilde{\alpha}_{\qq_1}^{*}a_0+\tilde{\beta}_{\bp} \tilde{\alpha}_{\qq_1}^{*}a^*_1\big|^2}{1+ e^2 |{A}|^2 \big| \tilde{\alpha}_{\bp}^{*}\tilde{\alpha}_{\qq_1}^{*}a_0+\tilde{\beta}_{\bp} \tilde{\alpha}_{\qq_1}^{*}a^*_1\big|^2}.
\eq%
We still need to evaluate the entropy of the electron-positron system, i.e. $\mathbb{\tilde{S}}_{e^-e^+}$, which is obtained by tracing out the photon modes of (\ref{DMfull}). In doing so, we find that the density matrix $\tilde{\varrho}_{e^-e^+}$ is not diagonal. After diagonalising it, only $2$ eigenvalues are non-null, so that the entropy of the pair is
\bq
\mathbb{\tilde{S}}_{e^-e^+} &=& -\sum_{i=0}^{1} \lambda''_i \log_2 \lambda''_i, \label{S-pair}
\eq with
\bq
\lambda''_0=\frac{1}{1+ e^2 |{A}|^2 \big| \tilde{\alpha}_{\bp}^{*}\tilde{\alpha}_{\qq_1}^{*}a_0+\tilde{\beta}_{\bp} \tilde{\alpha}_{\qq_1}^{*}a^*_1\big|^2},
\eq
\bq
\lambda''_1=\frac{e^2 |{A}|^2 \big| \tilde{\alpha}_{\bp}^{*}\tilde{\alpha}_{\qq_1}^{*}a_0+\tilde{\beta}_{\bp} \tilde{\alpha}_{\qq_1}^{*}a^*_1\big|^2}{1+ e^2 |{A}|^2 \big| \tilde{\alpha}_{\bp}^{*}\tilde{\alpha}_{\qq_1}^{*}a_0+\tilde{\beta}_{\bp} \tilde{\alpha}_{\qq_1}^{*}a^*_1\big|^2}.
\eq%
The fact that $\mathbb{\tilde{S}}_{e^-e^+} \neq 0$ indicates that the electron-positron system is no longer pure, that is, the pair is entangled with the emerging photon. Due to this, the mutual information, which quantifies the total (quantum $+$ classical) correlation between the pair, is expected to decrease, because the photon ``carries" part of the correlation.
Using (\ref{S-elec}), (\ref{S-pos}) and (\ref{S-pair}), we can compute the mutual information in the interacting case as
\begin{equation}
\tilde{\mathcal{I}}_{e^- e^+} = \tilde{\mathbb{S}}_{e^-} + \tilde{\mathbb{S}}_{e^+} - \tilde{\mathbb{S}}_{e^- e^+},
\end{equation}
and compare it with the free case in (\ref{Minfo-free}). We display our results in the graphs below. For the sake of simplicity we have consistently adopted $\qq = 0$ as its value does not affect qualitatively our conclusions. We have used $\frac{e^2}{4\pi}=\frac{1}{137}$.
\begin{figure}[H]
	\centering
	\includegraphics[scale=0.5]{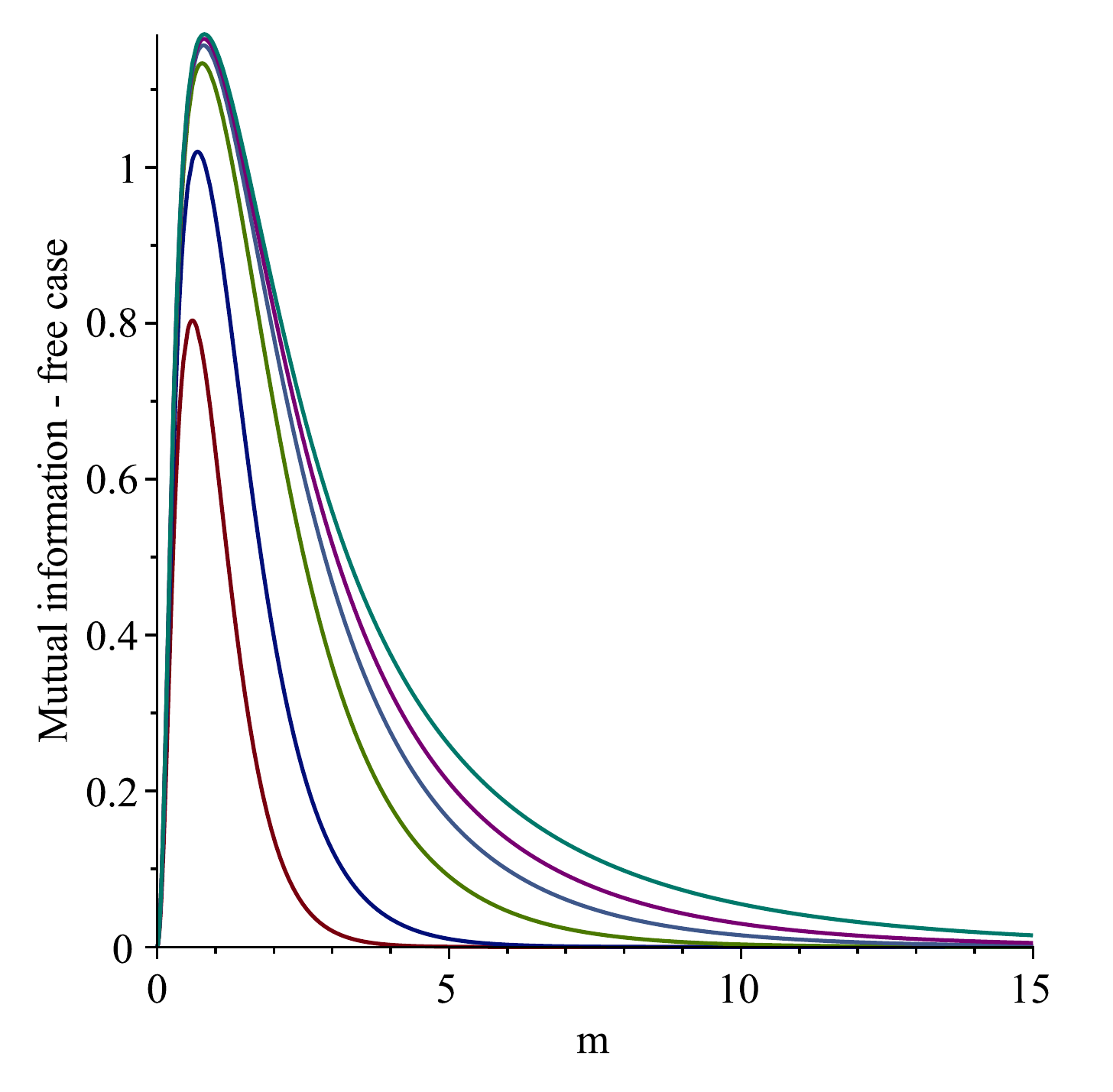}
	\caption{Mutual information between the electron and positron as a function of mass in the free case. We have used $|p|=1$ and $\epsilon=1$. The peaks increase with $\rho$ and saturate around $\rho\approx30$. Above we plotted the mutual informations corresponding to $\rho={3,5,10,15,20,30}$.}
	\label{G1}
\end{figure}
\begin{figure}[H]
	\centering
	\includegraphics[scale=0.5]{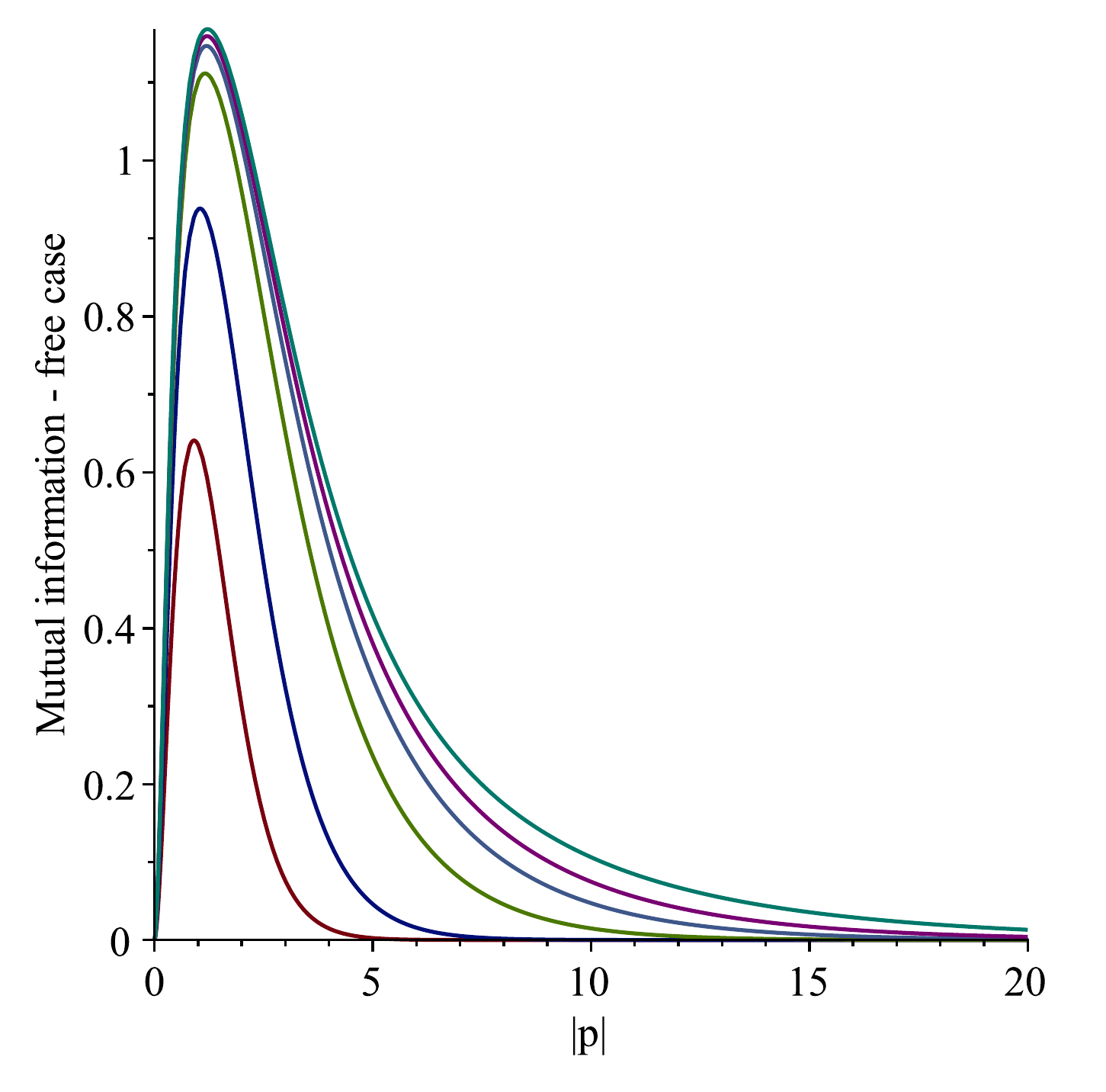}
	\caption{Mutual information between the electron and positron as a function of momentum in the free case. We have used $m=1$ and $\epsilon=1$. The peaks increase with $\rho$ and saturate around $\rho\approx30$ at an optimal momentum $|\text{p}|_{\text{optimal}}\approx1.2$. The curves correspond to $\rho={3,5,10,15,20,30}$. }
	\label{G2}
\end{figure}
\begin{figure}[H]
	\centering
	\includegraphics[scale=0.5]{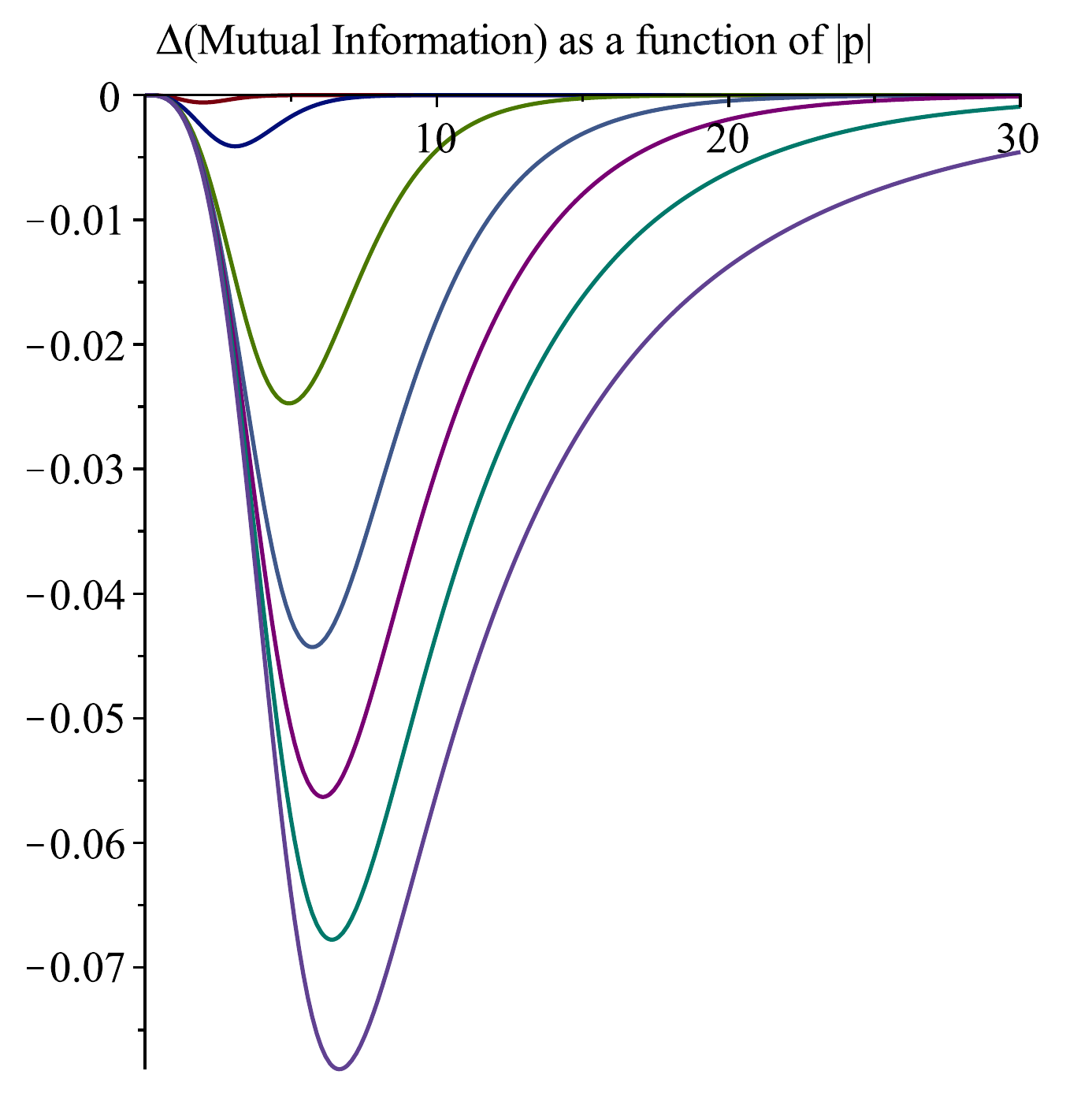}
	\caption{Decrease in the mutual information between the electron and positron due to the interaction ($\Delta\mathcal{I}_{e^-e^+}=\tilde{\mathcal{I}}_{e^-e^+}-\mathcal{I}^0_{e^-e^+}$). We have used $m=1$ and $\epsilon=1$. The curves correspond to $\rho={3,5,10,15,20,30}$ and $\rho=100$, where they saturate. Observe that the decrease is small up until around $|\text{p}|_{\text{optimal}}\approx1.2$ and for large momenta. The maximal mutual information loss for these parameters is about $6\%$.}
	\label{G3}
\end{figure}
\begin{figure}[H]
	\centering
	\includegraphics[scale=0.5]{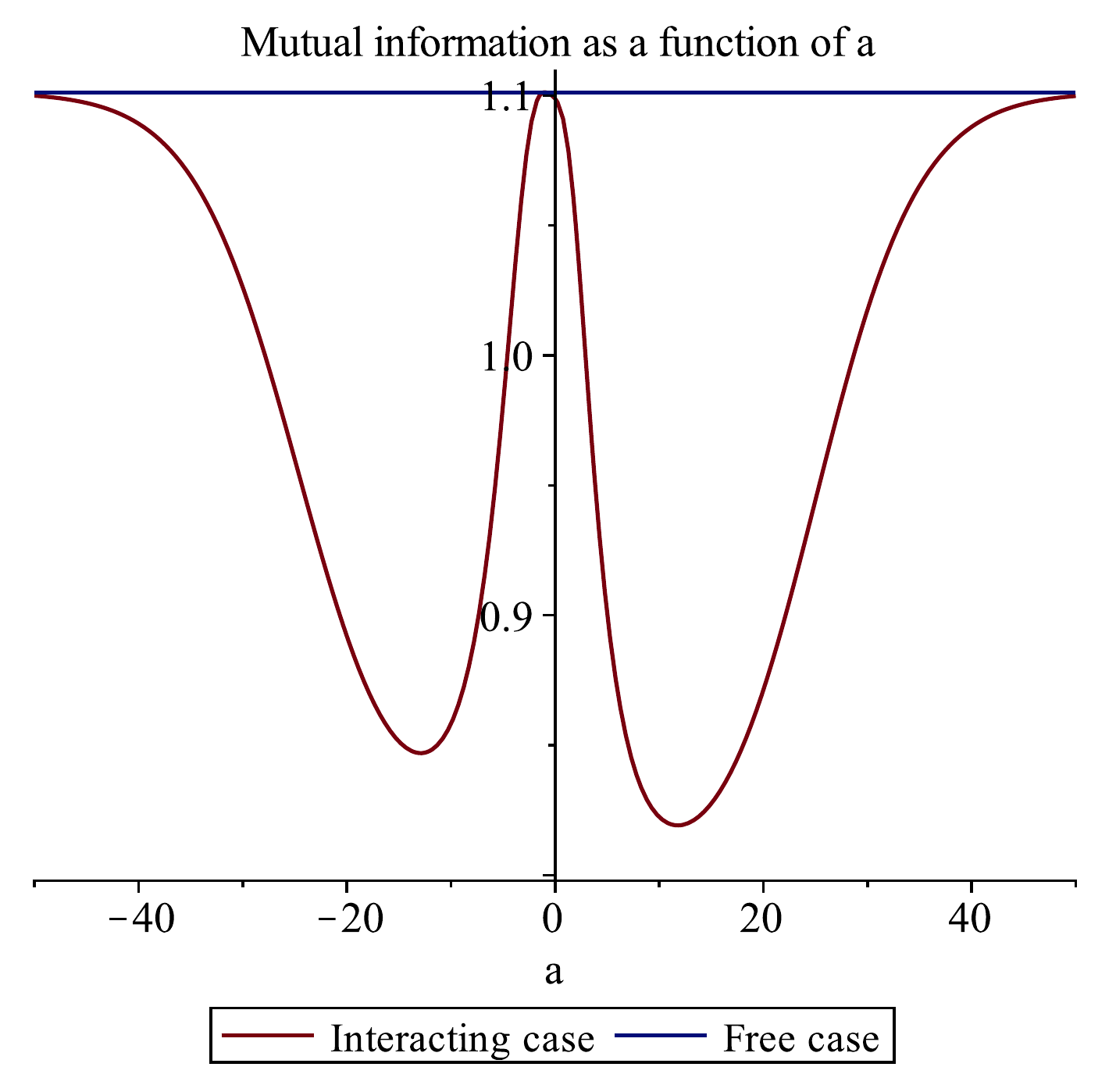}
	\caption{Mutual information of the electron-positron pair for $m=1$, $p=|\text{p}|_{\text{optimal}}$, $\epsilon=1$  and $\rho=30$. The mutual information for the free case corresponds to the flat line, while the interacting case is a function of $a$, which determines how the momentum $-\bp$ is distributed between the positron and photon. Notice that for $a=-1$ or $|a|>>1$, the mutual information for both cases coincide. Also, there are values of $a\approx\pm12$ which considerably decrease the mutual information in the interacting case.}
	\label{G4}
\end{figure}

\begin{figure}[H]
	\centering
	\includegraphics[scale=0.5]{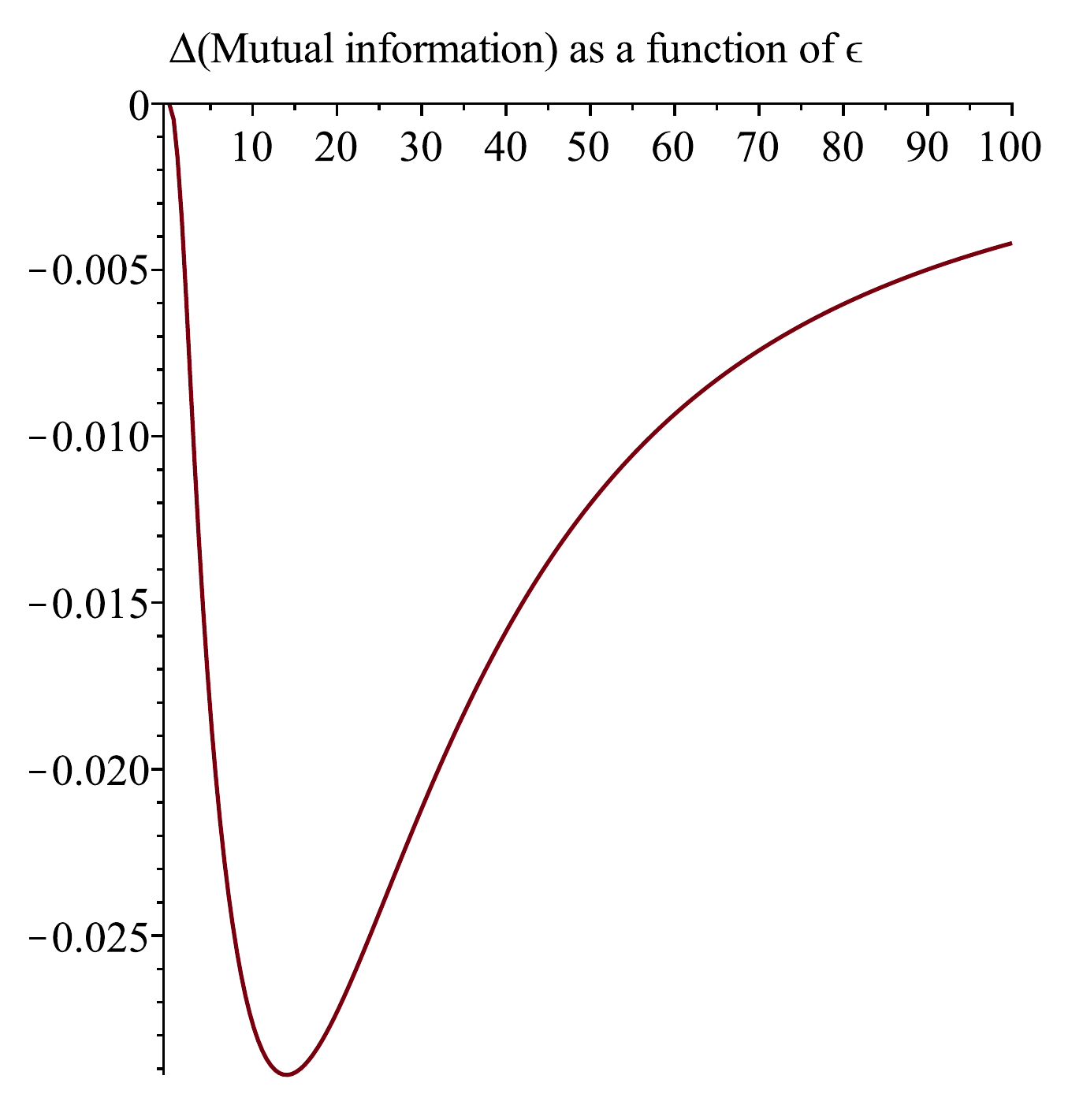}
	\caption{Variation of the mutual information of the electron-positron pair for $m=1$, $p=|\text{p}|_{\text{optimal}}$, and $\rho=30$ as a function of $\epsilon$.}
	\label{G5}
\end{figure}

\section{Discussion and conclusions}

It has been shown in \cite{FUENTES0},  for massive free scalar fields, and in \cite{FUENTES1}, for massive free fermionic fields,  that  information  about space-time evolution   is encoded in the entanglement between particles of opposite momenta as measured by the von Neumann entropy. For the Dirac vacuum in an expanding space-time, entangled fermion - anti-fermion pairs $(\bp, -\bp)$ are generated even in the absence of electromagnetic interaction for massive fermions. 
The entanglement entropy of the observed particles, say electrons, has distinguished features as compared with bosonic fields: whilst for fermions it peaks at an optimal frequency, for bosons it monotonically decreases. This somewhat shows that fermionic fields are more effective for extracting information about space-time evolution. 
Using such peculiar features of fermionic fields and within a  simple spatially flat, asymptotically minkowskian spacetime characterised by a rapidity $\rho$ and total volume $\epsilon$, it was possible to construct a protocol to estimate such cosmological parameters from the quantum correlations. Such protocols are described in \cite{FUENTES1}. In a few words, the rapidity estimation protocol is obtained from the optimal momentum, 
which is very sensitive to $\rho$ variations, whilst it changes very little with $\epsilon$. In addition, a lower bound for $\epsilon$ can be appraised by computing the maximum value of the quantum correlation via an optimal $|\bp|$ tuning as a function of $\epsilon$.  One of the purposes of our work is to study, to leading order in perturbation theory, how the electromagnetic interaction interferes with cosmological parameter assessment based on quantum correlations of modes of the pair electron-positron for an emerging electron of momentum $\bp$. 

Because a third particle is involved (photon) we employ the mutual information as a measurement of quantum correlations. It is adequate since it reduces to (twice) the von Neumann entanglement entropy of the electron (positron) in the limit where the coupling is zero. Figures 2 and 3 are just a restatement of the free case reported in \cite{FUENTES1}, save that we plotted the $e^- e^+$ mutual information showing that it peaks at optimal values of  the electron momentum $|\bp|$ for $\rho = 3, 5, 10, 15, 20$ and $30$. Figure 4 shows the variation 
$\Delta {\cal{I}} =\tilde{\cal{I}}_{e^- e^+}- {\cal{I}}_{e^- e^+}^0$ as a function of $|\bp|$ for the same values of $\rho$. Notice that it peaks at different values than figure 3 and vanishes for large $|\bp|$, showing that the decrease of mutual information is very small at the optimal momenta that maximise it. This decrease in the  $e^- e^+$ mutual information is due to the fact that the particles become correlated to the photon as well. The maximum decrease for the set of parameters we  adopted is around $6 \%$.  The behaviour of the $e^-e^+$ 
  mutual information as a function of $a$ (the parameter that characterises the momentum shared between the positron and the photon) is depicted in figure 5. Notice that it coincides with the free case (flat line) at $a=-1$, as it should, presents two minima at $a \approx \pm 12$, and approaches a value that coincides with the free case for $|a|>>1$. Moreover, from figure \ref{G4}, it is clear that multiple values of $a$ yield the same decrease in the mutual information, indicating that a $1$-$1$ correspondence between the mutual information and the outgoing momenta of the particles is not direct and should be taken into account in the evaluation of the cosmological parameters.
 Figure 6 shows the variation of the mutual information as a function of the volume of the universe $\epsilon$. Notice that it vanishes for large $\epsilon$ and peaks at a value of $\epsilon$ that corresponds to about 
$15\%$ of the mutual information of the free case for the same set of parameters.  

Finally it is worth mentioning that in \cite{HELDER1} we considered the influence of a self-interaction  $\lambda \phi^4$ for scalar fields $\phi$ on the entanglement entropy generation for a particular mode in an asymptotically free FRW space-time. It was shown that self-interaction enhances the entanglement entropy. Here, because of both the fermionic character  of the pair $e^-e^+$ and the interaction that involved a photon, we realise that some of the quantum correlation of the pair $e^-e^+$ is distributed to the photon meaning that in this case the interaction presents a deleterious effect.

Of course this analysis is qualitative and serves only to indicate  the percentage change in the quantum correlation due to the interaction. Such effects should be taken into consideration in the estimation protocols along with anisotropy effects and, last but not least, decoherence effects.
  
It would be interesting to construct more realistic models to take into account the effects above, as well as constructing appropriate detector models and detection processes for the interpretation of the measurements in the observation of (non-local) quantum correlations. In this sense, moving point-like detectors coupled to quantum fields have been considered to carry quantum information in space-time \cite{PL}. In \cite{AN1} finite-size detectors, i.e.
detectors with a position dependent coupling strength, are described and claimed 
not only to be more realistic but also to have the advantage of coupling to peaked distributions of modes.  This is important in the task of  evaluating correctly  the reality of these quantum correlations that define entanglement or even of effects that appear to not satisfy the causal propagation of signals as nicely discussed in \cite{BU}.

\section*{Acknowledgments}
\noindent
JBA [grant number 88881.188500/2018-01] and HASC thank CAPES for financial support. MS acknowledges FAPESP (2018/05948-6) and CNPq (303482/2017-6) for a research grant.

\section*{Appendix}
	
Let us explicitly compute the amplitude in equation (\ref{Gamma1230}),
$$
	\Gamma_{\bk_1,\bk_2,\bk_3} =  \langle \bk_3^\chi| \otimes \langle  \bk_2^{r'}, \bk_1^{r}| \Big( -i\int_{-\infty}^{\infty}\hat{H}_{\mathrm{int}} \,  d\eta  \Big) \, |0 \rangle^{D}_{\mathrm{in}} \otimes  | 0 \rangle^{\gamma}_{\mathrm{in}},
$$
where $\hat{H}_{\mathrm{int}} = -ie\int  d^3{\bf{x}} \, \sqrt{-g} \, \hat{\bar{\psi}} \gamma^{\mu} \hat{A}_{\mu} \hat{\psi} $. The mode expansion for the Dirac and Maxwell quantum fields are given by equations (\ref{eqnpsi}) and (\ref{eqnA}). Inserting the latter in the expression for $\Gamma_{\bk_1,\bk_2,\bk_3}$ yields
\bq
	&& \Gamma_{\bk_1,\bk_2,\bk_3} = -e\int d^4x \sqrt{-g} \sum_{s, s',\sigma} \int_{\bk_1',\bk_2',\bk_3'} \left[\frac{\mu_{\mathrm{in}}}{\omega_{\mathrm{in}}(k'_{1})} \right ]^{1/2} \times \nonumber \\ &&\times \left [\frac{\mu_{\mathrm{in}}}{\omega_{\mathrm{in}}(k'_{2})} \right ]^{1/2} \frac{e^{-i(\bold{k}'_{1} + \bold{k}'_{2} + \bold{k}'_{3})\cdot \bold{x}}}{\sqrt{2k'_{3}}}   
	   \bar{U}^{\mathrm{in}}_{\bk'_{1}, s}\gamma^{\mu}\epsilon_{\mu}^{\sigma *}(\bk'_{3})V^{\mathrm{in}}_{\bk'_{2},s} \nonumber \\
	 && \langle \bk^{r}_1 \bk^{r'}_2|\bk'^{s}_{1}\bk'^{s'}_{2}\rangle\langle \bk^{\chi}_3|\bk'^{\sigma}_{3}\rangle.
\eq
According to the normalization convention we have adopted, the scalar products read
\bq
&& \langle \bk^{r}_1 \bk^{r'}_2|\bk'^{s}_{1}\bk'^{s'}_{2}\rangle\langle \bk^{\chi}_3|\bk'^{\sigma}_{3}\rangle = \left (\frac{\omega_{\mathrm{in}}(k'_{1})}{\mu_{\mathrm{in}}} \delta^{(3)}(\bk'_{1}-\bk_{1}) \delta_{rs}  \right ) \nonumber \\
&& \left (\frac{\omega_{\mathrm{in}}(k'_{2})}{\mu_{\mathrm{in}}} \delta^{(3)}(\bk'_{2}-\bk_{2}) \delta_{r's'}  \right )  \left (2|k'_{3}| \delta^{(3)}(\bk'_{3}-\bk_{3}) \delta_{\chi \sigma}\right),\nonumber
\eq
which yields, after simple algebra, 
\bq 
\label{Gamma123II}
&&	\Gamma_{\bk_1,\bk_2,\bk_3} = -\frac{e}{(2\pi)^3(2\pi)^{\frac{3}{2}}} \int d^4x \sqrt{-g}  \left [\frac{\omega_{\mathrm{in}}(k_{1})}{\mu_{\mathrm{in}}} \right ]^{1/2} \nonumber \\ && \left [\frac{\omega_{\mathrm{in}}(k_{2})}{\mu_{\mathrm{in}}} \right ]^{1/2}  \sqrt{2k_{3}} \,\, 
 e^{-i(\bold{k}_{1} + \bold{k}_{2} + \bold{k}_{3})\cdot\bold{x}} \times \nonumber \\ &&  \times  \,\,  \bar{U}^{\mathrm{in}}_{\bk_{1}, r}\gamma^{\mu}\epsilon_{\mu}^{\chi *}(\bk_{3})V^{\mathrm{in}}_{\bk_{2},r'}.
\eq
To simplify the calculation, let us explicitly write the spinor functions as \cite{DUNCAN},
	\begin{align} \label{uvin}
	\begin{split}
	\bar{U}^{\mathrm{in}}_{\bk_{1}, r} &= -K_{\mathrm{in}}(k_{1}) (i\bold{\not{k}}_{1} + \mu_{\mathrm{in}}) e^{i\omega_{\mathrm{in}}(k_{1})\eta} \bar{U}^{\mathrm{in}}_{0, r}, \\
V^{\mathrm{in}}_{\bk_{2},r'} &= -K_{\mathrm{in}}(k_{2}) (i\bold{\not{k}}_{2} + \mu_{\mathrm{in}}) e^{i\omega_{\mathrm{in}}(k_{2})\eta} V^{\mathrm{in}}_{0,r'},
	\end{split}
	\end{align}
with $\bar{U}^{\mathrm{in}}_{0, r}$ and $V^{\mathrm{in}}_{0,r'}$ given by \cite{PESKIN}:
	\begin{align}
	\bar{U}^{\mathrm{in}}_{0, r} &= \sqrt{m} \begin{pmatrix}
	r^{\dagger} & r^{\dagger}
	\end{pmatrix}, \\
 V^{\mathrm{in}}_{0,r'} &= \sqrt{m} \begin{pmatrix}
	r'\\-r' 
	\end{pmatrix},
	\end{align}
and $r^{\dagger}=\begin{pmatrix}
	1 & 0
	\end{pmatrix}$ or $\begin{pmatrix}
	0 & 1
	\end{pmatrix}$ and $r'= \begin{pmatrix}
	1\\ 
	0
	\end{pmatrix}$ or $\begin{pmatrix}
	0\\ 
	1
	\end{pmatrix}$. 
For simplicity, consider a photon polarized in the z-direction. Thus, the polarization function $\epsilon_{\mu}^\chi$ is 
	\begin{align}
	\epsilon_{\mu}^\chi(\bk_3) = \frac{1}{\sqrt2}
	\begin{pmatrix}
	0\\ 
	1\\ 
	(-1)^{q}i\\ 
	0
	\end{pmatrix},
	\end{align}
	with $q = 1, 2$. Moreover, the product $\gamma^{\mu}\epsilon_{\mu}^q(\bk_3)$ is given by
\be
  \frac{1}{\sqrt2}  {\mathrm{AD}}\Big(1-(-1)^{q},1+(-1)^{q},-1+(-1)^{q},-1-(-1)^{q}\Big),\nonumber
\ee
where AD stands for anti-diagonal matrix going from the lower left corner to the upper right corner.	Thus we have,
\be
\bar{U}^{\mathrm{in}}_{0, r}\gamma^{\mu}\epsilon_{\mu}^{\chi *}(\bk_{3})V^{\mathrm{in}}_{0,r'} = \frac{4m}{\sqrt2}. 
\ee
Recalling that	 $\sqrt{-g} = C(\eta) = \left[1 + \epsilon(1 + \tanh(\rho \eta))\right]^2$, it can be shown that
\bq
&&	\int_{-\infty}^{+\infty} d\eta \left[1 + \epsilon(1 + \tanh(\rho \eta))\right]^2  e^{i\bar{\omega}\eta} = (1 + \epsilon)^2  (2\pi)^3 \delta(\bar{\omega}) + \nonumber \\ && + \left ( \frac{2\pi i}{\rho} \epsilon(1 + \epsilon) - \frac{\pi \epsilon \bar{\omega}}{\rho^2} \right )\frac{1}{\sinh(\frac{\pi \bar{\omega}}{2 \rho})}.
\eq
with $\bar{\omega}=\omega_{\mathrm{in}}(k_1) + \omega_{\mathrm{in}}(k_2)$.

Collecting all these results and plugging them back in (\ref{Gamma123II}) yields, factoring out $e \delta^3(\bk_1 + \bk_2+ \bk_3)$ according to our definition given in equation (\ref{Gamma1230}),
\bq
&&	A(k_1, k_2, k_3) = -\frac{4m}{(2\pi)^{\frac{3}{2}}} \left [\frac{\omega_{\mathrm{in}}(k_{1})}{\mu_{\mathrm{in}}} \right]^{1/2} \left[\frac{\omega_{\mathrm{in}}(k_{2})}{\mu_{\mathrm{in}}} \right]^{1/2} \nonumber \\ && \times \sqrt{k_{3}}  K_{\mathrm{in}}(k_{1}) K_{\mathrm{in}}(k_{2}) 
(i\bold{\not{k}}_{1} + \mu_{\mathrm{in}}) (i\bold{\not{k}}_{2} + \mu_{\mathrm{in}}) \times \nonumber \\ &&  \left [(1 + \epsilon)^2  (2\pi)^3 \delta(\bar{\omega})+\left( \frac{2\pi i}{\rho} \epsilon(1 + \epsilon) - \frac{\pi \epsilon \bar{\omega}}{\rho^2} \right)\frac{1}{\sinh(\frac{\pi \bar{\omega}}{2 \rho})}\right]\nonumber .
\eq

\end{document}